\documentclass[reprint, double column, superscriptaddress,prl, showkeys]{revtex4-1}

\usepackage{float}
\usepackage{graphicx,epsfig}
\usepackage{amssymb}
\usepackage{amsmath}
\usepackage{bm}
\usepackage{braket}
\newcommand* {\frack}[2]{{\textstyle\frac{#1}{#2}}}
\usepackage{textcomp}
\usepackage{color}
\usepackage{pgffor}

\usepackage{soul}

\usepackage[bookmarksnumbered,bookmarksopen]{hyperref}
\hypersetup{
   colorlinks=true,       
}

\usepackage{pdfpages}
\usepackage{BOONDOX-cal}

\makeatletter

\AtBeginDocument{\let\LS@rot\@undefined}
\makeatother

\newif\ifarXiv
\arXivtrue

\usepackage{soul}

\begin{document}
\setcounter{page}{1}

\title[]{Highly-Anisotropic Even-Denominator Fractional Quantum Hall State in an Orbitally-Coupled Half-Filled Landau Level}
\author{Chengyu \surname{Wang}}
\author{A. \surname{Gupta}}
\author{Y. J. \surname{Chung}}
\author{L. N. \surname{Pfeiffer}}
\author{K. W. \surname{West}}
\author{K. W. \surname{Baldwin}}
\affiliation{Department of Electrical and Computer Engineering, Princeton University, Princeton, New Jersey 08544, USA}
\author{R. \surname{Winkler}}
\affiliation{Department of Physics, Northern Illinois University, DeKalb, Illinois 60115, USA}
\author{M. \surname{Shayegan}}
\affiliation{Department of Electrical and Computer Engineering, Princeton University, Princeton, New Jersey 08544, USA}
\date{\today}

\begin{abstract}
The even-denominator fractional quantum Hall states (FQHSs) in half-filled Landau levels are generally believed to host non-Abelian quasiparticles and be of potential use in topological quantum computing. Of particular interest is the competition and interplay between the even-denominator FQHSs and other ground states, such as anisotropic phases and composite fermion Fermi seas.  Here we report the observation of an even-denominator fractional quantum Hall state with highly-anisotropic in-plane transport coefficients at Landau level filling factor $\nu=3/2$. We observe this state in an ultra-high-quality GaAs two-dimensional \textit{hole} system when a large in-plane magnetic field is applied. By increasing the in-plane field, we observe a sharp transition from an isotropic composite fermion Fermi sea to an anisotropic even-denominator FQHS. Our data and calculations suggest that a unique feature of two-dimensional holes, namely the coupling between heavy-hole and light-hole states, combines different orbital components in the wavefunction of one Landau level, and leads to the emergence of a highly-anisotropic even-denominator fractional quantum Hall state. Our results demonstrate that the GaAs two-dimensional hole system is a unique platform for the exploration of exotic, many-body ground states.

\end{abstract}

\maketitle  
When a strong perpendicular magnetic field ($B_{\perp}$) is applied to a two-dimensional electron system (2DES) at low temperatures, the thermal and kinetic energies are quenched as the electrons occupy the quantized Landau levels (LLs), and various many-body ground states emerge as the Coulomb energy dominates. Half-filled LLs are particularly interesting as they host a plethora of emergent, exotic quantum phases, including compressible composite fermion (CF) Fermi seas in the lowest orbital energy ($N=0$) LLs, e.g., at Landau level filling factors $\nu=1/2$ and $3/2$ in GaAs 2DESs \cite{Jain.Book.2007, Halperin.Jain.Book.2020}, and even-denominator fractional quantum Hall states (FQHSs) in the first-excited ($N=1$) LLs, e.g., at $\nu=5/2$ and $7/2$ \cite{Willett.PRL.1987, Pan.PRL.1999}. The even-denominator FQHSs have been of considerable, continued attention because they are generally expected to be Pfaffian states with non-Abelian statistics \cite{Haldane.PRL.1988, Moore.NPB.1991, Morf.PRL.1998, Nayak.RMP.2008}. FQHSs have also been reported at $\nu=3/2$ in different 2DESs. It was first observed in a wide GaAs quantum well with a bilayer-like charge distribution when the Fermi level ($E_F$) lies in the $N=0$ LL \cite{Suen.PRL.1994}. More recently, $\nu=3/2$ FQHSs were reported in different materials, such as ZnO \cite{Falson.Nat.Phy.2015}, AlAs \cite{Hossain.PRL.2018}, WSe$_2$ \cite{Shi.Nat.Nanotech.2020}, and bilayer graphene \cite{ Zibrov.Nature.2017, Li.Science.2017, Huang.PRX.2022}, when $E_F$ is in an $N=1$ LL. In contrast to these incompressible FQHSs in the $N=0$ and $N=1$ LLs, anisotropic phases emerge in the $N\geq 2$ LLs as the ground states at half fillings \cite{Koulakov.PRL.1996, Fogler.PRB.1996, Moessner.PRB.1996, Fradkin.PRB.1999, Fradkin.PRL.2000, Fradkin.ARCMP.2010, Lilly.PRL.1999, Du.SSC.1999, Fu.PRL.2020}. These are generally believed to be stripe phases which, at finite temperatures and in the presence of ubiquitous disorder, turn into nematic phases.

\begin{figure*}[t]
  \begin{center}
    \psfig{file=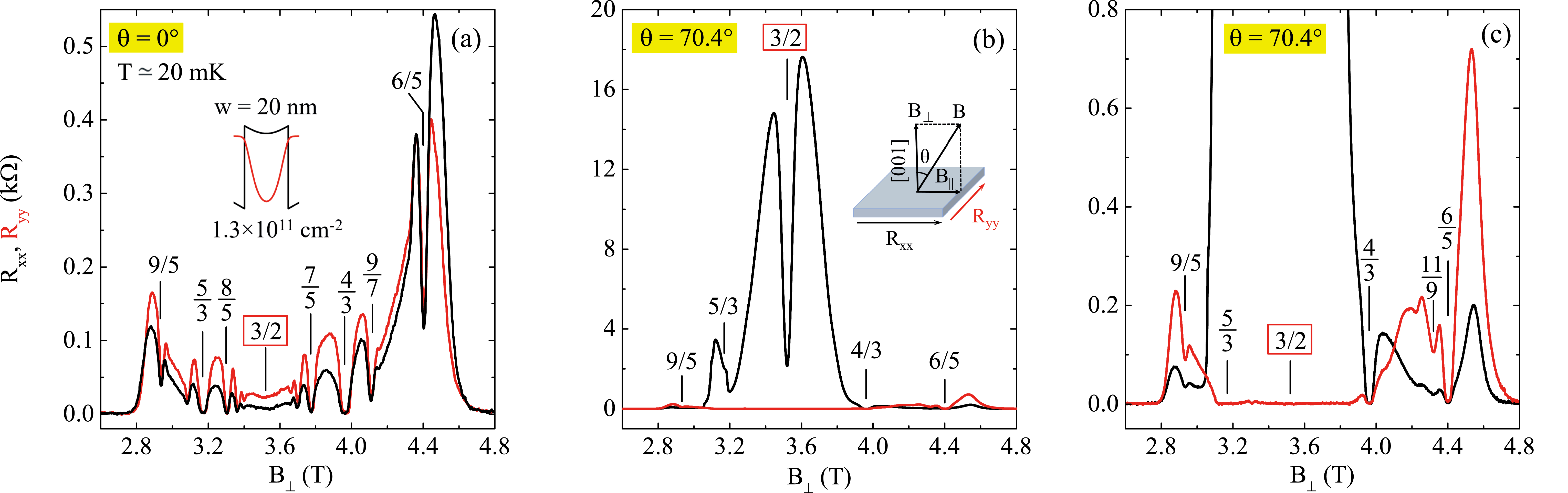, width=1\textwidth}
  \end{center}
  \caption{\label{base}
    Longitudinal resistances ($R_{xx}$ in black, and $R_{yy}$ in red) vs. perpendicular magnetic field $B_{\perp}$ at tilt angles: (a) $\theta=0^{\circ}$ and (b) $\theta=70.4^{\circ}$. (c) Enlarged version of (b). Inset in (a): The self-consistently-calculated hole charge distribution (red) and potential (black). We emphasize that our 2DHS is confined to a narrow quantum well and has a single-layer charge distribution. Inset in (b): A schematic of the experimental setup. The sample is mounted on a rotating stage to support \textit{in situ} tilt; $\theta$ is the angle between $B$ and $B_{\perp}$, and $R_{xx}$ and $R_{yy}$ denote resistances measured along and perpendicular to $B_{\parallel}$. }
  \label{fig:base}
\end{figure*}

As highlighted in Fig. 1, here we report a $B_{\parallel}$-induced, highly-anisotropic, even-denominator FQHS at $\nu=3/2$ in an ultra-high-mobility 2D \textit{hole} system (2DHS) confined to a GaAs quantum well \cite{Chung.PRM.2022, Footnote.3/4, Wang.PRL.2022, SM}. In a purely $B_{\perp}$, our 2DHS is nearly isotropic in the vicinity of $\nu=3/2$, and its in-plane longitudinal magneto-resistance traces ($R_{xx}$ and $R_{yy}$) exhibit smooth and shallow minima at $\nu=3/2$, flanked by the standard, odd-denominator FQHSs at $\nu=1+p/(2p\pm 1)$, such as $\nu=4/3, 7/5, 10/7, \ldots$, and $5/3, 8/5, 11/7, \ldots$ [Fig. 1(a)]. This is consistent with a compressible (Fermi sea) ground state of CFs at $\nu=3/2$, similar to what has been seen previously in very high-mobility GaAs 2DESs \cite{Jain.Book.2007, Halperin.Jain.Book.2020, Du.PRL.1995, Chung.NM.2021} and 2DHSs \cite{Liu.PRB.2016, Chung.PRM.2022}. When we tilt the sample in magnetic field to induce a large $B_{\parallel}$ component [Figs. 1(b,c)], our 2DHS breaks its rotational symmetry and becomes highly anisotropic near $\nu=3/2$, with $R_{xx}/R_{yy} \simeq 10^4$, and yet shows deep minima in both $R_{xx}$ and $R_{yy}$ at $\nu=3/2$ and $R_{xy} \simeq (2/3)(e^2/h)$ \cite{SM}. Note that $R_{xx}$ exceeds 10 k$\Omega$ near $\nu=3/2$ whereas $R_{yy}\lesssim 1$ $\Omega$. 

Figure 1(b,c) traces are truly intriguing. On the one hand, the deep minima in both $R_{xx}$ and $R_{yy}$ at $\nu=3/2$, and an $R_{xy}\simeq (2/3)(e^2/h)$ are consistent with a developing FQHS at this filling. On the other hand, the enormous transport anisotropy near $\nu=3/2$ is reminiscent of the $B_{\parallel}$-induced anisotropic states that are observed in GaAs 2DESs in the $N=1$ LL, e.g. at $\nu=5/2$, when the sample is tilted to a large angle $\theta$ \cite{Lilly.PRL.1999.tilt, Pan.PRL.1999.tilt, Xia.PRL.2010, Liu.PRB.2013, Friess.PRL.2014, Shi.PRB.2015}. Note, however, that the anisotropic states observed at $\nu=5/2$ at large $\theta$ replace the FQHS which is present at small $\theta$ \cite{Lilly.PRL.1999.tilt, Pan.PRL.1999.tilt, Xia.PRL.2010, Liu.PRB.2013, Friess.PRL.2014, Shi.PRB.2015}. In contrast, in our 2DHS, at $\nu=3/2$, an even-denominator FQHS with highly-anisotropic transport coefficients replaces a compressible (CF Fermi sea) ground state at small $\theta$ \cite{Samkharadze.NP.2016, Schreiber.NC.2018, Falson.SA.2018, Footnote.phases}. Also remarkable is the presence of numerous FQHSs at odd-denominator filings such as $4/3$, $6/5$, $9/5$, and $11/9$ in Fig. 1(c) at large $\theta$. These FQHSs are generally seen in 2DESs in an $N=0$ LL \cite{Pan.PRB.2008, Footnote.second.LL, Kumar.PRL.2010}. The even-denominator FQHS at $\nu=3/2$ and its anisotropic behavior, however, are telltale signs that $E_F$ lies in an $N>0$ LL. In this Letter we present the evolution of this very unusual, anisotropic FQHS at $\nu=3/2$ with temperature and $B_{\parallel}$.  Our data and calculations suggest that a unique feature of 2D holes, namely the coupling between heavy-hole (HH) and light-hole (LH) states, combines $N=0$ and $N>0$ components in the wavefunction of one LL, and leads to the emergence of the anisotropic FQHS at $\nu=3/2$.

The ultra-high-quality 2DHS studied here is confined to a $20$-nm-wide GaAs quantum well grown on a GaAs (001) substrate. The 2DHS has a density of $1.3\times 10^{11}$ cm$^{-2}$ and a low-temperature ($0.3$ K) record-high mobility of $5.8\times 10^{6}$ cm$^2$/Vs \cite{Chung.PRM.2022}. We performed our experiments on a $4\times 4$ mm$^2$ van der Pauw geometry sample, with alloyed In:Zn contacts at the four corners and side midpoints. The sample was then mounted on a rotating stage and cooled in a dilution refrigerator. We measured the longitudinal resistances, $R_{xx}$ and $R_{yy}$, along and perpendicular to $B_{\parallel}$ (see Fig. 1(b) inset), using the lock-in amplifier technique.

\begin{figure*} [t]
  \begin{center}
    \psfig{file=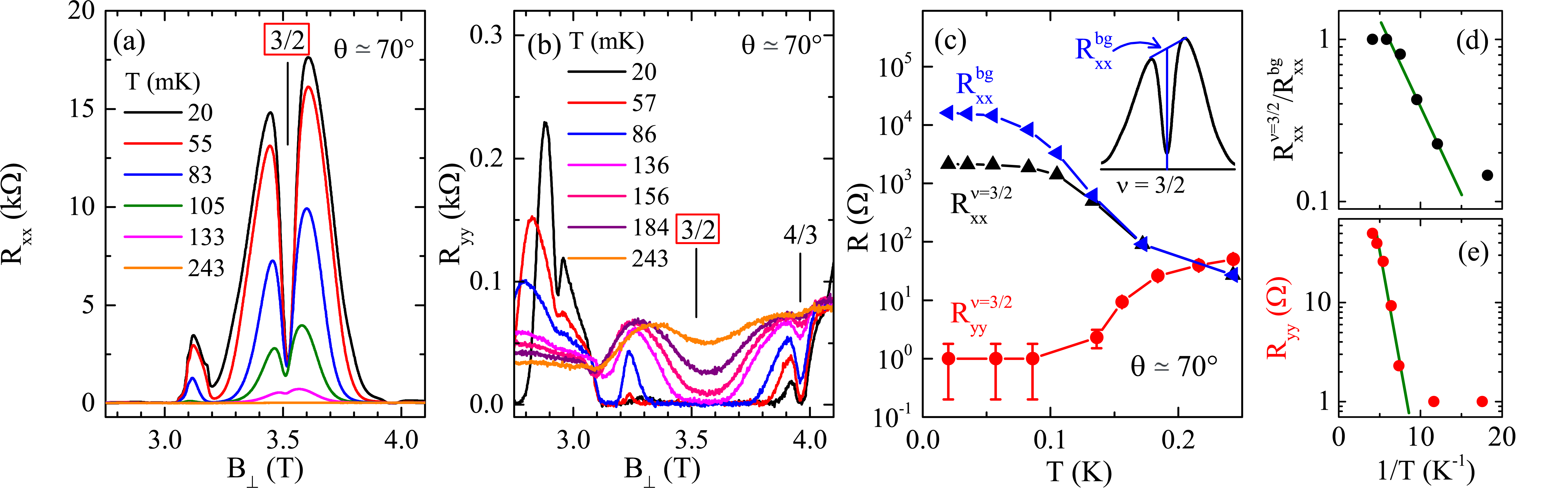, width=1\textwidth}
  \end{center}
  \caption{\label{Tdep} 
     Temperature dependence of $R_{xx}$ and $R_{yy}$. (a, b) $R_{xx}$ and $R_{yy}$ vs. $B_{\perp}$ traces near $\nu=3/2$ taken at $\theta \simeq 70^{\circ}$ and at different temperatures. (c) $R^{\nu=3/2}_{xx}$, $R^{\nu=3/2}_{yy}$, and $R^{bg}_{xx}$ vs. $T$. Inset shows an $R_{xx}$ vs. $B_{\perp}$ trace near $\nu=3/2$, showing how $R_{xx}^{bg}$ at $\nu=3/2$ is defined. (d, e) Arrhenius plots of $R^{\nu=3/2}_{xx}/R^{bg}_{xx}$ and $R_{yy}$ vs. $1/T$. The lines are linear fits to the data at intermediate temperatures, and their slopes give energy gaps $\simeq 0.5$ and $\simeq 2.1$ K for (d) and (e) respectively. 
}
  \label{fig:Tdep}
\end{figure*}

In Figs. 2(a, b), we show $R_{xx}$ and $R_{yy}$ vs. $B_{\perp}$ traces taken at $\theta \simeq 70^{\circ}$ at different temperatures. With decreasing temperature, $R_{xx}$ near $\nu=3/2$ grows whereas $R_{yy}$ becomes smaller and approaches zero. Figure 2(c) presents $R^{\nu=3/2}_{xx}$, $R^{\nu=3/2}_{yy}$, and the background resistance $R^{bg}_{xx}$ \cite{Footnote.Rbg} on the flanks of $\nu=3/2$ vs. $T$. The data show that the 2DHS is highly anisotropic at low temperatures but becomes nearly isotropic at $T\gtrsim 0.2$ K. To quantify the strength of the $\nu=3/2$ resistance minimum and extract an estimate for the FQHS energy gap, we make Arrhenius plots of $R_{yy}$ and the ratio $R^{\nu=3/2}_{xx}/R^{bg}_{xx}$ as a function of $1/T$ [Figs. 2(d, e)]. Note that, at $\nu=3/2$, $R_{xx}$ does not decrease as temperature is lowered, likely because of the strong rise in $R_{xx}$ on the flanks of $\nu=3/2$. We therefore plot $R^{\nu=3/2}_{xx} / R^{bg}_{xx}$ to obtain an estimate for the $\nu=3/2$ FQHS energy gap \cite{Mallett.PRB.1988}. The lines in Figs. 2(d, e) are linear fits to the data points at intermediate temperatures, and their slopes give an estimate of the $\nu=3/2$ FQHS gap, $\sim 1$ K.

Measuring the Hall resistance, $R_{xy}$, for a highly-anisotropic 2D system is very challenging and seldom reported \cite{Xia.Nat.Phys.2011, Hossain.PRL.2018}. Near $\nu=3/2$ our 2DHS becomes very resistive along the hard-axis direction ($R_{xx} \simeq 20$ k$\Omega$), which inevitably induces non-uniformities in the current distribution in the sample and can also cause severe $R_{xx}$ mixing to $R_{xy}$. The effect of $R_{xx}$ mixing can be minimized by averaging $R_{xy}$ taken for opposite polarities of $B$ \cite{Sajoto.PRL.1993, Goldman.PRL.1993}. We did such averaging and, although we do not see a quantized Hall plateau at $\nu=3/2$, we find that $R_{xy}$ approaches $(2/3)(h/e^2)$ to within $1\%$; see Supplemental Material (SM) \cite{SM} for details. The fact that $R_{xy}$ does not exhibit a well-developed plateau is consistent with the deep, but non-zero, $R_{xx}$ minimum.

\begin{figure}[t!]
  \begin{center}
    \psfig{file=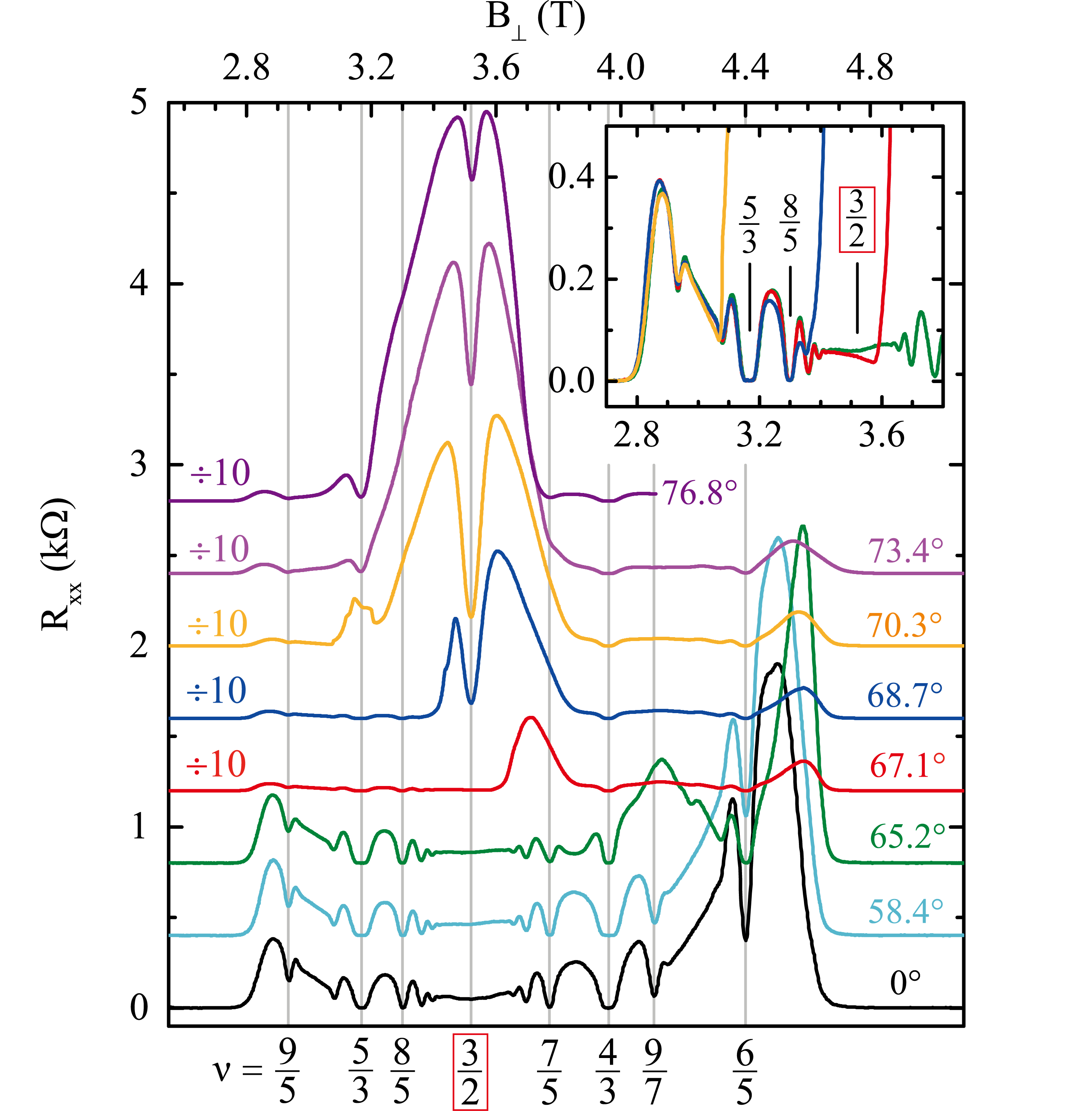, width=0.5\textwidth}
  \end{center}
  \caption{\label{tilt} 
   Tilt-angle dependence of $R_{xx}$ is shown vs. $\nu$ at $20$ mK. Each trace is vertically shifted by $0.4$ k$\Omega$ for clarity. The heights of some traces are divided by $10$, as marked on the left. Inset: An enlarged version of $R_{xx}$ vs. $B_{\perp}$ at $\theta=65.2^{\circ}$, $67.1^{\circ}$, $68.7^{\circ}$, and $70.3^{\circ}$.}
  \label{fig:tilt}
\end{figure}

We now turn to the evolution of the different phases at and near $\nu=3/2$ with tilt-angle $\theta$. Figure 3 shows $R_{xx}$ vs. $\nu$ traces at different $\theta$ \cite{Footnote.Config}. With increasing $\theta$, the FQHSs go through transitions, starting at the lowest $\nu$ (highest $B_{\perp}$). For example, the $R_{xx}$ minimum at $\nu=9/7$ turns into a maximum at $\theta=65.2^{\circ}$ and disappears at higher $\theta$. At larger $\theta$, $R_{xx}$ rises very abruptly above a $B_{\perp}$ which depends on $\theta$ (see Fig. 3 inset), and attains very high values in a range of fields near $\nu=3/2$ (main Fig. 3). The ground state at $\nu=3/2$ exhibits an abrupt transition from a CF Fermi sea at $\theta \leq 65.2^{\circ}$ to a highly-anisotropic FQHS when $\theta \geq 68.7^{\circ}$. The strongest $\nu=3/2$ FQHS is seen at $\theta \simeq 70^{\circ}$. With further increase in $\theta$, $R_{xx}$ remains large, and the $\nu=3/2$ $R_{xx}$ minimum gradually gets weaker, while all the FQHSs between $\nu=4/3$ and 5/3 disappear. We also observe qualitatively similar transitions, including the sudden appearance of a $\nu=3/2$ FQHS, in another GaAs 2DHS with a higher density ($\simeq 2.2\times10^{11}$ cm$^{-2}$); however, the transitions occur at smaller $\theta$ ($\simeq 60^{\circ}$ instead of $\simeq 69^{\circ}$); see SM \cite{SM, Footnote.crossing}. 

The evolution of the FQHSs near $\nu=3/2$ presented in Fig. 3 is consistent with a crossing of the 2DHS LLs induced by increasing $B_{\parallel}$. Such a crossing has been previously reported \cite{Liu.PRB.2016, Zhang.PRB.2017, Footnote.differences}, and can be understood based on the energy ($E$) vs. $B_{\perp}$ LL diagram shown in Fig. 4. This LL diagram is from calculations performed for our 2DHS in a purely $B_{\perp}$ using the multiband envelope function approximation (EFA) and the $8 \times 8$ Kane Hamiltonian \cite{Winkler.Book.2003}. Because of the strong HH-LH coupling, the LLs of GaAs 2DHSs are highly nonlinear and show numerous crossings \cite{Winkler.Book.2003}. There are two crossings between the top two LLs, one between the $\boldsymbol \alpha$ level (in blue) and $\boldsymbol \beta$ level (in green) at very low $B_{\perp}$ ($\simeq 1$ T), and another between the $\boldsymbol \beta$ level and $\boldsymbol \gamma$ level (in red) at higher $B_{\perp}$ ($\simeq 18$ T). These two crossings are by themselves very interesting and lead to many exotic interaction phenomena \cite{Liu.PRB.2014, Liu.PRB.2015, Liu.PRL.2016, Ma.PRL.2022} near $\nu=1/2$ and 1. 

The relevant crossing in our study is between the $\boldsymbol \alpha$ and $\boldsymbol \gamma$ levels \cite{Liu.PRB.2016}. Note that in Fig. 4 this crossing occurs at $\nu < 3/2$. While we cannot perform calculations for a 2DHS in tilted fields because of computational challenges, we expect that the addition of $B_{\parallel}$ would increase the effective Zeeman energy and the separation between the $\boldsymbol \alpha$ and $\boldsymbol \beta$ levels. This implies that the crossing between $\boldsymbol \alpha$ and $\boldsymbol \gamma$ would move to higher fillings, eventually reaching $\nu=3/2$ and even higher $\nu$. Our data support this scenario. At $\theta=0^{\circ}$, $E_F$ at $\nu=3/2$ lies in the $\boldsymbol \alpha$ level (Fig. 4). This is essentially an $N=0$ LL, consistent with our observation of a compressible CF Fermi sea ground state at $\nu=3/2$, flanked by the standard, odd-denominator FQHSs at $\nu=1+p/(2p\pm 1)$ [Figs. 1(a) and 3]. As we increase $\theta$ and, therefore, $B_{\parallel}$, the $B_{\perp}$ position of the crossing moves toward lower $B_{\perp}$. At sufficiently large $\theta$, $E_F$ at $\nu=3/2$ moves to the other side of the crossing, and lies in the $\boldsymbol \gamma$ level. This is manifested by an extremely abrupt rise of $R_{xx}$ at $\nu=3/2$ as a function of $B_{\parallel}$ (Fig. 4 inset). We elaborate on the above crossing in the SM \cite{SM} based on our measured strengths of the different FQHSs near $\nu=3/2$, and show that it occurs in a very narrow range of $B_{\parallel}$ near 8 T (see Fig. S5 of SM \cite{SM}).

Next we discuss how HH-LH coupling in 2DHSs combines $N=0$ and $N>0$ components in the wavefunction of one LL, which can explain the unusual behavior observed in our experiments.
In the EFA, LLs in 2D systems can
be described using two sets of dimensionless ladder operators $a,
a^\dagger$ and $b, b^\dagger$, corresponding to two harmonic
oscillators \cite{Suzuki.PRB.1974, MacDonald.1995,
Winkler.Book.2003}, see also SM \cite{SM}.  The single-particle
Hamiltonian $H$ only depends on the $a$ oscillators; the $a$
oscillators thus define the energy spectrum of $H$ as a function of
$B_\perp$. The $b$ operators commute with $H$, and represent an oscillator with frequency zero that describes the degeneracy of the LLs.

In the EFA applied to the $s$-like electrons in the conduction band
of GaAs, the wavefunctions have two spinor components corresponding
to a spin $s = \pm 1/2$.  However, ignoring the weak spin-orbit
coupling in electron systems, the spin is decoupled from the orbital
motion, and eigenstates of $H$ have only one nonzero spinor
component $s$.  Therefore, the quantum number $N = 0, 1, 2, \ldots$
that labels the eigenstates $\ket{N}$ of the $a$ oscillator can also
label the eigenstates of $H$ with energies $E_N = \hbar \omega_c (N
+ \frac{1}{2}) \pm \Delta_\mathrm{Z}$ and Zeeman energy
$\Delta_\mathrm{Z}$.

The $p$-like holes in the valence band require four spinor
components representing $s= \pm 3/2$ (HH) and $\pm 1/2$ (LH) \cite{Footnote.spinor}. The hole eigenstates of $H$ are
generally a linear combination of HH and LH states.  However, to a
good approximation, the eigenstates of $H$ remain eigenstates of
total angular momentum $j = l + s$ (axial approximation
\cite{Suzuki.PRB.1974, Winkler.Book.2003}), where $l=xp_y - yp_x = \hbar \left(a^\dagger a - b^\dagger b\right)$ is the orbital angular momentum.  The eigenstates of $H$ become
\begin{equation}
  \label{eq:ax-state}
  \psi_\mathcal{N} = \sum_s \ket{N = \mathcal{N} - s - \frack{3}{2}}
  \, \xi^\mathcal{N}_s u_s
\end{equation}
with $\ket{N} = 0$ when $N < 0$.  Here, $u_s$ are the base spinors
with spin $s$ and $\xi^\mathcal{N}_s$ gives the relative weights of
the HH and LH spinors.  We ignore in Eq.\ (\ref{eq:ax-state}) the
$b$ oscillators that are not relevant for the present discussion.
$\mathcal{N} = 0, 1, 2, \ldots$ labels the energies of the LLs.  In
contrast to 2DESs, $N$ is not a good quantum number for hole LLs
because of the HH-LH mixing that is parameterized via the
coefficient $\xi^\mathcal{N}_s$ \cite{Footnote.2DES.mixing}. 

Note that in Fig. 4 the $\boldsymbol \alpha$ level is a pure HH LL
($s=-3/2$, $\mathcal{N}=0$) containing only $N=0$, consistent with
Eq.\ (\ref{eq:ax-state}).  The $\boldsymbol \gamma$ level with
$\mathcal{N}=3$, on the other hand, is strongly affected by HH-LH
coupling, which makes this level a mixture of different oscillator
states $\ket{N}$ associated with the different spinor components
representing the HH and LH parts of its wavefunction.  The
decomposition of the $\boldsymbol \gamma$ level yields approximately
49\% $s=+3/2$ with $N=0$, 45\% $s=+1/2$ with $N=1$, and 6\% $s=-3/2$
with $N=3$. This can explain why the traces we observe at and near
$\nu=3/2$ at large angles, e.g., at $\theta \simeq 70^{\circ}$, are
intriguingly unique as they exhibit features normally seen in an
\textit{electron} system in LLs with different $N$: $N=0$ (numerous
odd-denominator FQHSs at 6/5, 11/9, etc.), $N=1$ (an
even-denominator FQHS), and $N>1$ LLs (states with extremely
anisotropic transport coefficients). It is also worth emphasizing
that, while the anisotropic $\nu=3/2$ FQHS in our sample abruptly
appears after the LL crossing at $B_{\parallel} \simeq 8$ T, it
persists well past the crossing, even when $B_{\parallel} \simeq 15$
T.

\begin{figure}[t!]
  \begin{center}
    \psfig{file=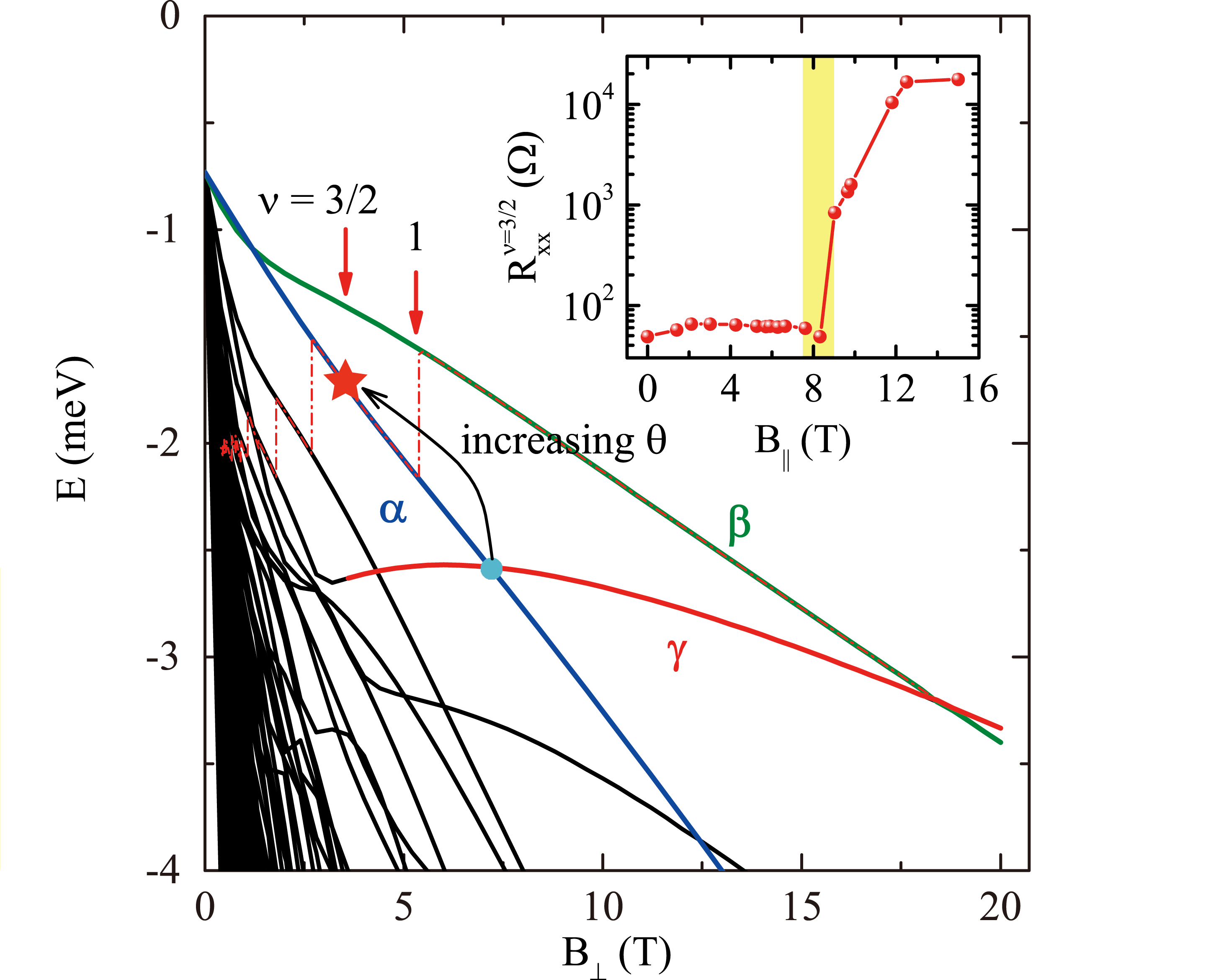, width=0.5\textwidth}
  \end{center}
  \caption{\label{LL} 
   Calculated LL diagram at $\theta=0^{\circ}$. The red dash-dotted line traces the Fermi energy. The position of $E_F$ at $\nu=3/2$ is marked with a star. Inset: $R_{xx}$ at $\nu=3/2$ vs. $B_{\parallel}$.
   }
  \label{fig:LL}
\end{figure}

We emphasize that the above calculations are performed in purely perpendicular magnetic fields and should be interpreted cautiously when tilted fields are involved. In large $B_{\parallel}$, the HH-LH coupling may be different. It is unlikely that the reduced symmetry in tilted fields would reduce the amount of HH-LH coupling experienced by the $\boldsymbol \gamma$ level. It is indeed more likely that the eigenstates in a tilted field would become yet more complicated linear combinations of different oscillator states $\ket{N}$. Note that such combinations are unique in GaAs 2DHSs, and cannot be achieved in GaAs 2DESs. Our results presented here demonstrate the power of applying $B_{||}$ as a knob to engineer the properties of 2D hole LLs and create exotic, new ground states. They also highlight the need for calculations of 2DHS LLs in titled magnetic fields; such calculations, while computationally very challenging, would be very helpful in understanding the physics of the observed ground states.

\begin{acknowledgments}

We acknowledge support by the National Science Foundation (NSF) (Grants No. DMR 2104771 and No. ECCS 1906253) for measurements, the U.S. Department of Energy (DOE) Basic Energy Sciences (Grant No. DEFG02-00-ER45841) for sample characterization, and the Eric and Wendy Schmidt Transformative Technology Fund and the Gordon and Betty Moore Foundation's EPiQS Initiative (Grant No. GBMF9615 to L. N. P.) for sample fabrication. Our measurements were partly performed at the National High Magnetic Field Laboratory (NHMFL), which is supported by the NSF Cooperative Agreement No. DMR 1644779, by the State of Florida, and by the DOE. This research is funded in part by QuantEmX grant from Institute for Complex Adaptive Matter and the Gordon and Betty Moore Foundation through Grant No. GBMF9616 to C. W., A. G., and M. S. We thank L. Jiao and T. Murphy at NHMFL for technical assistance, and Jainendra K. Jain and Ajit C. Balram for illuminating discussions.

\end{acknowledgments}



\section*{Supplementary material}

# Supplemental Material to "Highly-Anisotropic Even-Denominator Fractional Quantum Hall State in an Orbitally-Coupled Half-Filled Landau Level"

Chengyu Wang,[1] A. Gupta,[1] Y. J. Chung,[1] L. N. Pfeiffer,[1] K. W. West,[1] K. W. Baldwin,[1] R. Winkler,[2] and M. Shayegan[1]

[1]*Department of Electrical and Computer Engineering, Princeton University, Princeton, New Jersey 08544, USA*

[2]*Department of Physics, Northern Illinois University, DeKalb, Illinois 60115, USA*

(Dated: July 7, 2023)

# I. ADDITIONAL TRANSPORT DATA

In Fig. S1, we present magneto-resistance $R_{xx}$ (in black) and $R_{yy}$ (in red) vs. perpendicular magnetic field $B_\perp$ of the two-dimensional hole system (2DHS) studied in the main text. The traces are measured under purely perpendicular magnetic fields ($\theta = 0°$) at a temperature of $\simeq 20$ mK. $B_\perp$ positions of typical filling factors are marked with vertical solid lines. The 2DHS shows nearly isotropic transport coefficients near $\nu = 1/2$, $3/2$ and also at $\nu \geq 4$. Numerous conventional, odd-denominator, fractional quantum Hall states (FQHSs) are seen on the flanks of $\nu = 1/2$ and $3/2$, attesting to the exceptionally high quality of our 2DHS [1]. A deep $R_{xx}$ minimum is observed in both $R_{xx}$ and $R_{yy}$ at an even-denominator filling $\nu = 3/4$, indicating a very unusual FQHS, which has been discussed in detail in Ref. [2]. Near $\nu = 5/2$ and $7/2$, however, the 2DHS becomes anisotropic with $R_{yy} \simeq 5R_{xx}$; qualitatively similar to what has been reported previously for 2DHSs [3–6]. Also, there is only a very weak hint of a $\nu = 5/2$ FQHS: A weak resistance minimum is seen at $\nu = 5/2$ in $R_{yy}$ while there is no clear minimum in $R_{xx}$ [see Fig. S1 (inset)], similar to what was observed in a higher-density 2DHS [4].

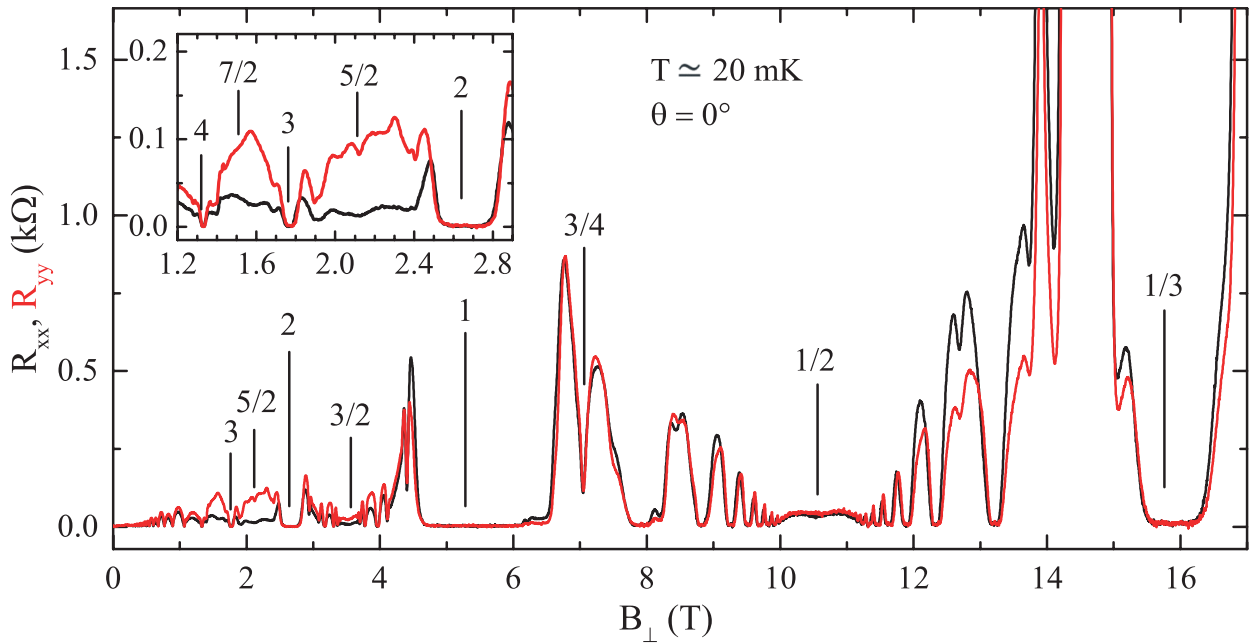

Fig. S1. Full-field $R_{xx}$ and $R_{yy}$ vs. $B_\perp$ traces at $\theta = 0°$. $B_\perp$ positions of typical filling factors are marked with vertical solid lines. Inset: Enlarged version near $\nu = 5/2$ and $7/2$.

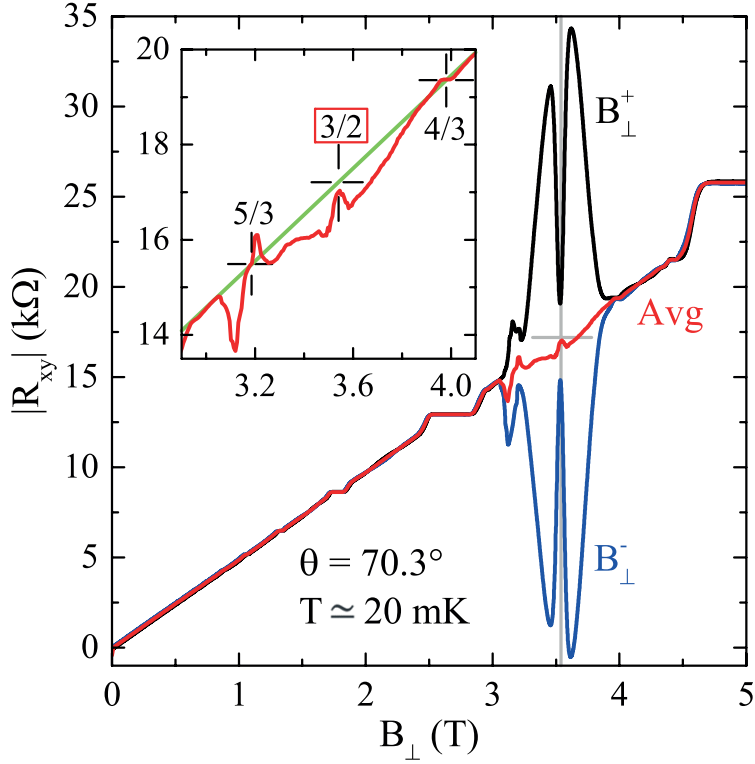

Fig. S2. Hall traces at $\theta = 70.3°$ taken at $T \simeq 20$ mK for two opposite polarities of magnetic field ($B_\perp^+$ in black, and $B_\perp^-$ in blue). The red trace shows the average of the $B_\perp^+$ and $B_\perp^-$ traces. Inset: An enlarged version of the Hall trace near $\nu = 3/2$. Expected $B_\perp$ positions and $R_{xy}$ of typical filling factors are marked. A straight line with a slope deduced from the low-$B_\perp$ Hall trace is shown in light green.

Figure S2 shows Hall traces for our 2DHS at $\theta = 70.3°$ taken for opposite polarities of magnetic field ($B_\perp^+$ and $B_\perp^-$). The traces were taken by passing current between two opposite corner contacts and measuring the Hall voltage developed across the other two corner contacts. Note that the $|R_{xy}|$ traces for $B_\perp^+$ and $B_\perp^-$ overlap in the whole field range except near $\nu = 3/2$, where they significantly deviate from the linear trend, and in opposite directions. This is because near $\nu = 3/2$ our 2DHS becomes highly anisotropic and very resistive along the hard-axis direction ($R_{xx} \simeq 20$ k$\Omega$), which inevitably induces non-uniformities in the current distribution in the sample and can also cause severe $R_{xx}$ mixing to the Hall traces. The effect of $R_{xx}$ mixing can be minimized by averaging $R_{xy}$ taken for opposite polarities of $B$ [7, 8]. We did such averaging and, although we do not see a quantized Hall plateau at $\nu = 3/2$, we find that $R_{xy}$ approaches $(2/3)(h/e^2)$ to within 1%

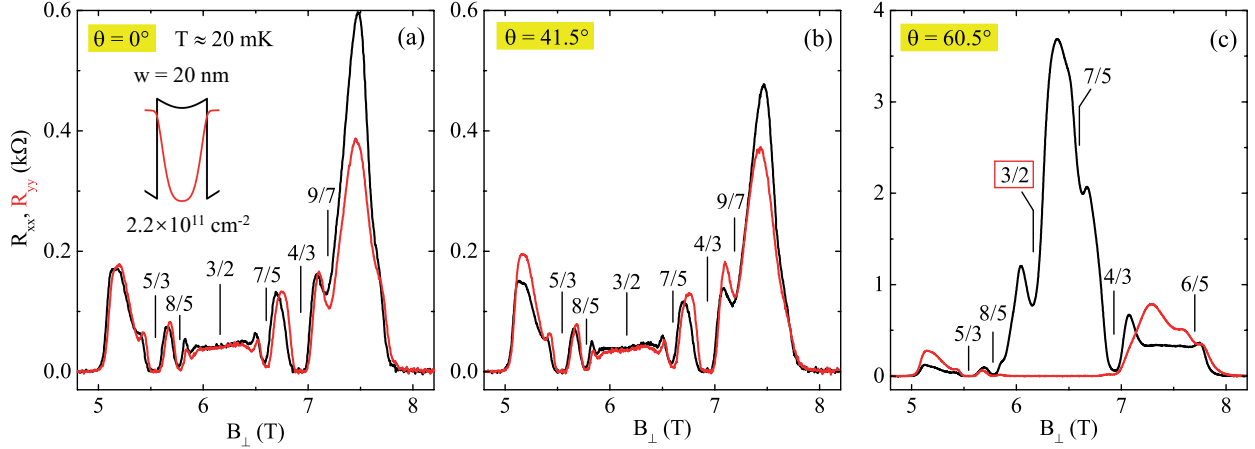

Fig. S3. $R_{xx}$ and $R_{yy}$ vs. $B_\perp$ traces for a sample with a higher 2D hole density ($\simeq 2.2 \times 10^{11}$ cm$^{-2}$) at tilt angles (a) $\theta = 0°$, (b) $\theta = 41.5°$, and (c) $\theta = 60.5°$. Inset in (a): The self-consistently-calculated hole charge distribution (red) and potential (black). We emphasize that our 2DHS is confined to a narrow quantum well and has a single-layer charge distribution.

(see Fig. S2 and its inset). The fact that $R_{xy}$ does not exhibit a well-developed plateau is consistent with the deep, but non-zero, $R_{xx}$ minimum. We emphasize that measuring the Hall resistance for a highly-anisotropic 2D system is very challenging and seldom reported [9, 10].

## II. HIGHLY-ANISOTROPIC $\nu = 3/2$ FRACTIONAL QUANTUM HALL STATE IN A HIGHER DENSITY SAMPLE

We measured another sample with a higher 2D hole density ($\simeq 2.2 \times 10^{11}$ cm$^{-2}$), which also shows a transition from a compressible (Fermi sea) ground state to a highly-anisotropic FQHS at $\nu = 3/2$. In Fig. S3, we present $R_{xx}$ and $R_{yy}$ vs. $B_\perp$ traces for this sample at three different tilt angles. At $\theta = 0°$ and $41.5°$, the 2DHS is isotropic in the vicinity of $\nu = 3/2$, with no FQHS features at $\nu = 3/2$. Numerous standard, odd-denominator FQHSs are observed at $\nu = 1 + p/(2p \pm 1)$, such as $\nu = 4/3, 7/5, ...,$ and $5/3, 8/5, ...$ [Figs. S3 (a, b)]. These attest to the exceptionally high quality of the sample. At $\theta = 60.5°$, the 2DHS becomes highly anisotropic between $\nu = 8/5$ and $4/3$, with $R_{xx}$ exceeding 3 k$\Omega$ and $R_{yy} \lesssim 1$ $\Omega$, and yet shows a minimum in both $R_{xx}$ and $R_{yy}$ at $\nu = 3/2$. The data are overall very similar to what we observe in the lower-density sample, except that the anisotropic $\nu = 3/2$

4

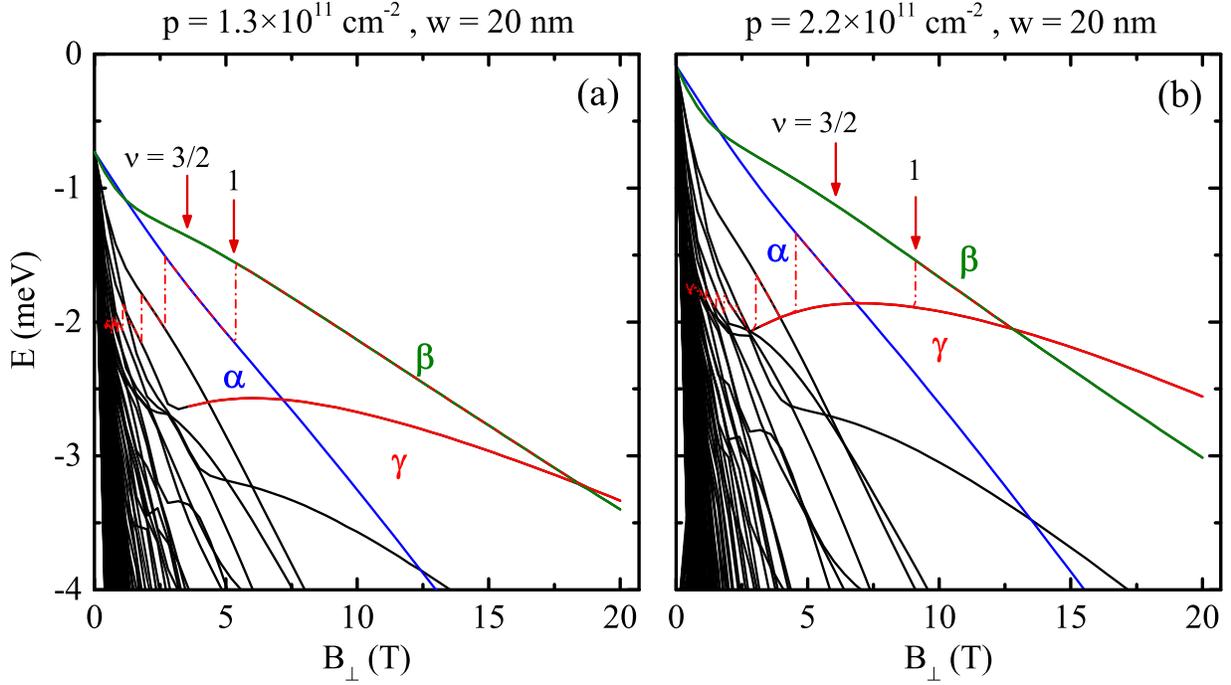

Fig. S4. Calculated energy $E$ vs. $B_\perp$ LL diagrams at $\theta = 0°$ for the two samples in our experiments: (a) $p = 1.3 \times 10^{11}$ cm$^{-2}$, adapted from Fig. 4 of the main text, (b) $p = 2.2 \times 10^{11}$ cm$^{-2}$.

FQHS emerges at a smaller tilt angle ($\theta \simeq 60°$ instead of $\theta \simeq 70°$). This emergence at a smaller $\theta$ is indeed what we would expect as higher density generally moves the crossing between the $\boldsymbol{\alpha}$ and $\boldsymbol{\gamma}$ levels in Fig. 4 towards a larger filling factor.

Our Landau level calculations indeed support our above conjecture. In Fig. S4, we present calculated energy $E$ vs. $B_\perp$ LL diagrams at $\theta = 0°$ for the two samples in our experiments. For the lower-density sample ($p = 1.3 \times 10^{11}$), the crossing between the $\boldsymbol{\alpha}$ and $\boldsymbol{\gamma}$ levels occurs at $\nu < 1$. For the higher-density sample ($p = 2.2 \times 10^{11}$), the crossing occurs at a larger filling factor ($1 < \nu < 3/2$). Therefore, it is reasonable that a smaller tilt angle is needed for the higher-density sample to tune the crossing between the $\boldsymbol{\alpha}$ and $\boldsymbol{\gamma}$ levels to $\nu = 3/2$.

The fact that we see qualitatively similar phenomena in two samples with very different densities (a factor of 1.7) indicates that the highly-anisotropic FQHS at $\nu = 3/2$ in large $B_\parallel$ is a robust phenomenon in very-high-quality 2DHSs.

## III. PREVIOUS STUDIES OF GROUND STATES OF GAAS TWO-DIMENSIONAL HOLE SYSTEMS AT $\nu = 3/2$ IN TILTED MAGNETIC FIELDS

The ground states of GaAs 2DHSs at $\nu = 3/2$ have been studied in tilted magnetic fields [12, 13]. In this section, we compare our sample and data with the previous results.

In Ref. [12], several GaAs 2DHSs confined to *wide* quantum wells were investigated in large tilt angles. By increasing the tilt angle, a transition from an isotropic composite fermion (CF) Fermi sea to an anisotropic phase with *no FQHE features* was observed at $\nu = 3/2$. In contrast, we observe an anisotropic FQHS at $\nu = 3/2$ at large tilt angles in our 2DHSs confined to *narrow* quantum wells. The absence of the FQHS at $\nu = 3/2$ in Ref. [12] could be attributed to two factors:

(i) Quantum well width. The samples in Ref. [12] are wide-well (30 nm and 35 nm) 2DHSs with *bilayer* charge distributions (see Fig. 1(b) and Fig. 3(b) inset in Ref. [12]), while our samples are narrow-well (20 nm) 2DHSs with *single-layer* charge distributions (see Fig. S3(a) inset, and Fig. 1(a) inset in the main text). A large in-plane magnetic field is known to make the charge distribution even more bilayer-like [14]. The wide quantum wells and the ensuing bilayer charge distributions could modify the interaction between holes at $\nu = 3/2$, and lead to a different ground state.

(ii) Quality of the 2DHSs. The samples in our study have exceptionally high quality thanks to the recent improvement in molecular beam epitaxy [1, 15]. The sample whose results are presented in the main text has a record-high mobility of $5.8 \times 10^6$ cm$^2$/Vs at 0.3 K. The high quality is also evinced by the observation of numerous Jain-sequence FQHSs surrounding $\nu = 3/2$, e.g., at $\nu = 4/3, 7/5, ...,$ up to $13/9$, and $5/3, 8/5, ...,$ up to $14/9$ [see Fig. 1(a) of the main text]. The higher-density sample in Fig. S3 also exhibits a very high quality as we discussed in the previous section. In comparison, the samples in Ref. [12] have much lower mobility ($\simeq 10^6$ cm$^2$/Vs), and show fewer FQHSs on the flanks of $\nu = 3/2$. It is likely that the lower sample quality leads to the absence of a FQHS at $\nu = 3/2$.

The 2DHSs in Ref. [13] have quantum well widths (17.5 nm and 20 nm) similar to our samples, but much lower mobility (0.6 - 1.7×$10^6$ cm$^2$/Vs). However, no FQHS was observed at $\nu = 3/2$ in those samples under a large in-plane magnetic field. This implies that the high quality of our 2DHSs is essential for the observation of the anisotropic FQHS at $\nu = 3/2$.

It is also noteworthy that in Ref. [12], when the charge distribution was made slightly

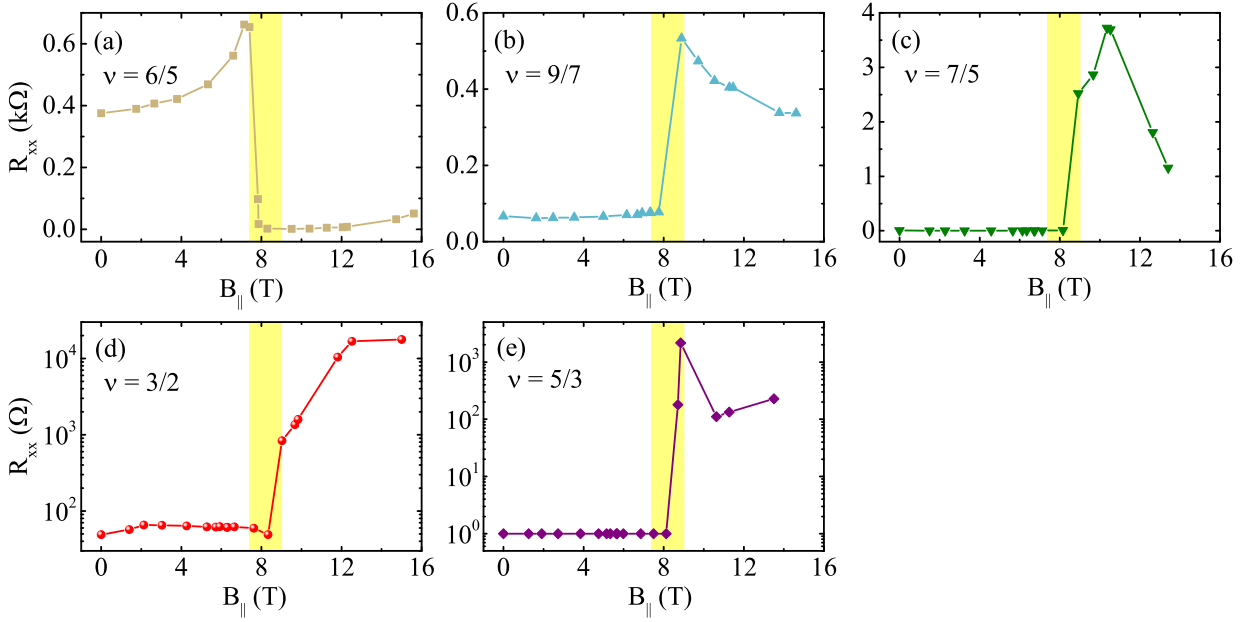

Fig. S5. $R_{xx}$ vs. $B_\parallel$ at different $\nu$: (a) 6/5, (b) 9/7, (c) 7/5, (d) 3/2, (e) 5/3. The yellow band shows the narrow range of $B_\parallel$ from 7.4 T to 9 T where sharp transitions at fractional fillings near $\nu = 3/2$ occur.

*asymmetric*, the anisotropic phase was replaced by a FQHS at $\nu = 3/2$ in a narrow range of $\theta$. However, the data is very limited as a clear FQHS feature at $\nu = 3/2$ only appears in a single trace at one tilt-angle of $\theta = 61°$. It is not clear why a small charge asymmetry would lead to a different ground state, i.e., to an even-denominator FQHS without significant transport anisotropy, at $\nu = 3/2$. In our work, we do not investigate the charge distribution symmetry because our samples are not fitted with gates. The role of charge asymmetry on the ground states of GaAs 2DHSs at $\nu = 3/2$ is indeed very interesting and requires more experimental and theoretical efforts in the future.

## IV. EVIDENCE FOR LANDAU LEVEL CROSSING

In Fig. S5, we show the evolutions of several FQHSs with increasing $B_\parallel$ for the sample studied in the main text (density $\simeq 1.3 \times 10^{11}$ cm$^{-2}$). The data were taken at $T \simeq 20$ mK. At a certain $\nu$, a larger $R_{xx}$ generally indicates a weaker FQHS. Sharp transitions are seen in the field range of 7.4 T $< B_\parallel <$ 9 T, as highlighted by a yellow-color band. However, these

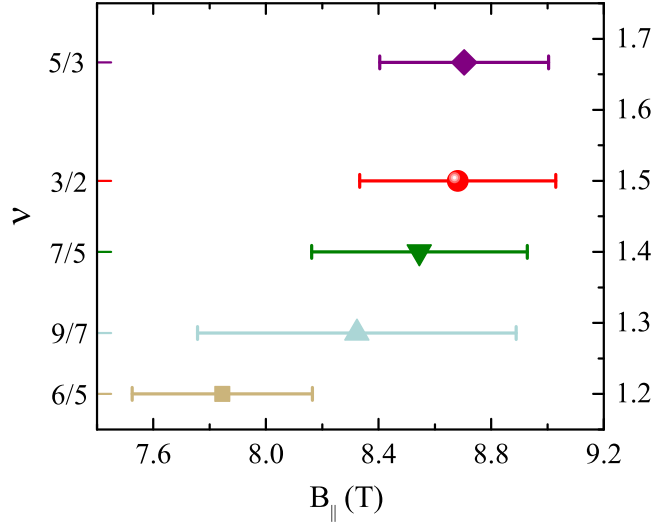

Fig. S6. Filling factor $\nu$ vs. the $B_\parallel$ position of the transitions at different $\nu$.

transitions exhibit qualitatively different behaviors. At $\nu = 6/5$, the FQHS, as signaled by the depth of its $R_{xx}$ minima, is weak at small $B_\parallel$, but becomes significantly stronger immediately after the transition, followed by a gradual weakening with further increase in $B_\parallel$ [Fig. S5(a)]. At $\nu = 9/7$, $7/5$ and $5/3$, in contrast, the FQHSs are strong at small $B_\parallel$, but suddenly get weaker and even disappear at the transition, and then reappear at larger $B_\parallel$ [Figs. S5(b, c, e)]. The situation at $\nu = 3/2$ [Fig. S5(d)] is more subtle because of the simultaneous emergence of the FQHS and anisotropy. On the one hand, the 2DHS becomes more anisotropic and thus shows an increasing $R_{xx}$ with increasing $B_\parallel$, starting from the transition. On the other hand, a FQHS at $\nu = 3/2$, evinced by a deep $R_{xx}$ minimum, appears simultaneously when the 2DHS becomes anisotropic near $\nu = 3/2$. The $B_\parallel$ dependence of $R_{xx}$ at $\nu = 3/2$ seen in Fig. S5(d) is dominated by the onset of the anisotropic phase and its extremely large resistance along the hard-axis direction, namely $R_{xx}$.

In the following, we discuss the evidence for a LL crossing near $\nu = 3/2$ in our 2DHS. ($i$) As seen in Fig. S6, all the transitions near $\nu = 3/2$ occur within a narrow $B_\parallel$ range, strongly suggesting that they have the same origin. The transitions occur at higher $B_\parallel$ for higher $\nu$, consistent with the expected evolution of the crossing of the $\boldsymbol{\alpha}$ and $\boldsymbol{\gamma}$ levels with $\theta$ (or $B_\parallel$); see LL diagram of Fig. 4 of the main text and the related discussion. ($ii$) The onset of the anisotropic phase near $\nu = 3/2$ in $B_\perp$ is very abrupt: As seen in Fig. 3 of the main text

and its inset, at large $\theta$, starting from the lower $B_\perp$ side where $E_F$ still lies in the $\boldsymbol{\alpha}$ level, the $R_{xx}$ vs. $B_\perp$ traces largely follow the $\theta = 0°$ trace. At certain $B_\perp$, when $E_F$ crosses to $\boldsymbol{\gamma}$, $R_{xx}$ suddenly increases, deviating dramatically from the $\theta = 0°$ trace. Moreover, as we increase $\theta$, the onset of the anisotropic phase moves towards lower $B_\perp$ as the crossing occurs at lower $B_\perp$. (*iii*) In GaAs 2DESs, even-denominator FQHSs and $B_\parallel$-induced anisotropic phases at half-fillings are usually observed in the $N = 1$ LLs [9, 11, 16–22]. In our 2DHS, the $\boldsymbol{\alpha}$ level is purely an $N = 0$ LL, but the $\boldsymbol{\gamma}$ level is more complex (see Section IV for a detailed discussion). To conclude, the $B_\parallel$-induced, anisotropic, even-denominator FQHS at $\nu = 3/2$ appears only at large $\theta$ when $E_F$ lies in the $\boldsymbol{\gamma}$ level.

## V. HEAVY-HOLE LIGHT-HOLE COUPLING IN GAAS TWO-DIMENSIONAL HOLE SYSTEMS

We cannot perform calculations for a 2DHS in tilted fields because of computational challenges. Here we present a more detailed discussion of the calculations of GaAs hole LLs in a perpendicular magnetic field $B_\perp$ that provide guidance regarding the nature of the LLs in the tilted-field experiments.

Holes in the topmost valence band have an effective spin 3/2. In a quantum well the four-fold degeneracy is lifted and we get heavy holes (HH) corresponding to spin $z$ component $s = \pm 3/2$ and light holes (LH) with $s = \pm 1/2$. For in-plane wave vector $k_\parallel = 0$ and $B_\perp = 0$ the eigenstates are purely HH or LH, whereas $k_\parallel \neq 0$ or $B_\perp \neq 0$ give rise to HH-LH coupling. (For conceptual clarity we restrict the discussion to the so-called axial approximation for 2D holes. We also restrict ourselves to the subspace representing the topmost valence band, and ignore excited subbands. A more complete discussion can be found in Ref. [23].) The eigenfunctions of the Hamiltonian are then four-component spinors representing $s = +3/2, +1/2, -1/2$ and $-3/2$.

The LLs in 2D systems can be described using two sets of dimensionless ladder operators [23–25]:

$$a = \frac{-i}{\sqrt{2}}\left(\frac{\bar{w}}{2\ell} + 2\ell\frac{\partial}{\partial w}\right), \qquad a^\dagger = \frac{i}{\sqrt{2}}\left(\frac{w}{2\ell} - 2\ell\frac{\partial}{\partial \bar{w}}\right), \qquad (1a)$$

$$b = \frac{1}{\sqrt{2}}\left(\frac{w}{2\ell} + 2\ell\frac{\partial}{\partial \bar{w}}\right), \qquad b^\dagger = \frac{1}{\sqrt{2}}\left(\frac{\bar{w}}{2\ell} - 2\ell\frac{\partial}{\partial w}\right), \qquad (1b)$$

corresponding to two harmonic oscillators. Here $w = x + iy$, the bar indicates complex

9

conjugation, and $\ell = \sqrt{\hbar/(eB_\perp)}$ is the magnetic length. The single-particle Hamiltonian $H$ only depends on the $a$ oscillators; the $a$ oscillators thus define the energy spectrum of $H$ as a function of $B_\perp$. The $b$ oscillators commute with $H$, and represent an oscillator with frequency zero that describes the degeneracy of the LLs.

When $B_\perp \neq 0$, the total angular momentum perpendicular to the 2D plane $j = l + s$ remains a good quantum number, where:

$$l = xp_y - yp_x = \hbar \left( a^\dagger a - b^\dagger b \right) \tag{2}$$

is the orbital angular momentum [23, 24]. The LLs with quantum number $\mathcal{N} = 0, 1, 2, \ldots$ can thus be written as [23]:

$$\psi_{\mathcal{N},M}(\boldsymbol{r}) = \sum_s |N = \mathcal{N} - s - \tfrac{3}{2}\rangle \, |M\rangle \, \xi_s^{\mathcal{N}}(z) \, u_s(\boldsymbol{r}). \tag{3}$$

Here $|N\rangle$ with $N = 0, 1, 2, \ldots$ denotes the eigenstates of the $a$ oscillator with $|N\rangle = 0$ when $N < 0$, $|M\rangle$ with $M = 0, 1, 2, \ldots$ are the eigenstates of the $b$ oscillator, $\xi_s^{\mathcal{N}}(z)$ expresses the weight of each spinor component $s$, and $u_s(\boldsymbol{r})$ are band-edge Bloch functions that act as the basis vectors for the spinor components of the multi-spinor wave functions $\psi_{\mathcal{N},M}(\boldsymbol{r})$.

The index $N$ representing the states $|N\rangle$ is an important quantum number in the fractional quantum Hall regime. In 2DESs, each LL is characterized by one value of $N$, and LLs with different $N$ host different ground states. According to Eq. (3), the LLs formed by 2D holes are linear combinations of different Landau oscillators $|N\rangle$ representing the four spinor components.

In Fig. S4 (a), we present the calculated $E$ vs. $B_\perp$ LL diagram for our 2DHS. Because of the HH-LH coupling, LLs are highly nonlinear as a function of $B_\perp$ and show numerous crossings and anti-crossings. The most relevant hole LLs in our study near $\nu = 3/2$ are the $\boldsymbol{\alpha}$ and $\boldsymbol{\gamma}$ levels. The $\boldsymbol{\alpha}$ level with $\mathcal{N} = 0$ is special as it remains pure HH ($s = -3/2$ with $N = 0$) consistent with Eq. (3); it does not undergo HH-LH coupling [26]. This is the reason why this level yields an almost perfectly straight line as a function of $B_\perp$. The $\boldsymbol{\gamma}$ level with $\mathcal{N} = 3$, on the other hand, is severely affected by HH-LH coupling. This level is already past an anti-crossing (at $\simeq 3$ T). The decomposition of the wave function yields approximately 49% $s = +3/2$ with $N = 0$, 45% $s = +1/2$ with $N = 1$, and 6% $s = -3/2$ with $N = 3$.

## VI. PREVIOUS EXPERIMENTAL RESULTS ON ANISOTROPIC FRACTIONAL QUANTUM HALL STATES

Highly anisotropic FQHSs have been rarely observed. They have been reported in GaAs 2DESs at $\nu = 5/2$ and $7/3$, in the $N = 1$ LL, when a relatively small $B_\parallel$ is applied [9, 20]. In these cases, the "hard-axis" ($R_{xx}$) is along the direction of $B_\parallel$, similar to what we observe. However, these FQHSs are isotropic at $\theta = 0°$ and exhibit a gradual isotropic to anisotropic transition with increasing $B_\parallel$, and eventually disappear at large $B_\parallel$, giving way to an anisotropic state (with no FQHS features) which is generally interpreted to be a stripe or nematic phase. This is very different from our data: We observe an isotropic compressible (Fermi sea) ground state at $\theta = 0°$, and this state abruptly turns into a highly-anisotropic FQHS at very large $B_\parallel$.

An anisotropic even-denominator FQHS has also been reported at $\nu = 3/2$ when $E_F$ lies in an $N = 1$ LL in a 2DES confined to an *AlAs* quantum well [10]. In this case, there is no $B_\parallel$ (*i.e.*, $\theta = 0°$), and the anisotropic 3/2 FQHS is attributed to the anisotropy of the in-plane effective mass in the AlAs 2DES [10]. Also noteworthy is that the data in Ref. [10] show only a robust and quantized $R_{xy}$ plateau at $\nu = 3/2$ but no hint of an $R_{xx}$ minimum.

## VII. THEORETICAL INTERPRETATIONS OF ANISOTROPIC FRACTIONAL QUANTUM HALL STATES

Theoretically, anisotropic FQHSs in tilted magnetic fields have been studied using effective field theory [27] and microscopic models [28, 29]. In Ref. [27], it is proposed that both $R_{xx}$ and $R_{yy}$ should eventually vanish at the lowest temperatures. This was observed in Ref. [20] for the $\nu = 5/2$ FQHS in tilted fields, but was not seen in Ref. [9] for the $\nu = 7/3$ FQHS. In our data, at the lowest temperatures, $R_{yy}^{\nu=3/2}$ becomes immeasurably small, while $R_{xx}^{\nu=3/2}$ saturates at a high value, likely because of the very resistive background.

Yang [29] discussed the conditions for the coexistence of nematic and FQHS phases, namely for the stability of a nematic FQHS. Yang showed that, when the Coulomb energy exceeds the LL separation, the ensuing LL mixing can potentially stabilize nematic FQHSs. This might explain why ,in a large $B_\parallel$, the 3/2 FQHS in our 2DHS, where LL mixing is severe because of large 2D hole effective mass, the anisotropic FQHS at $\nu = 3/2$ is robust,




\begin{thebibliography}{99}
\bibitem{Footnote.Rbg} We define background resistance  near $\nu=3/2$ ($R^{bg}_{xx}$) by connecting a straight line between the two $R_{xx}$ peaks on the two sides of $\nu=3/2$ and picking the resistance value at the field position of $\nu=3/2$.
\bibitem{Footnote.Config} In Fig. 3 $R_{xx}$ is measured with a different contact configuration compared to $R_{xx}$ in Figs. 1 and 2.
\end{thebibliography}
\end{document}